\documentclass[aps,prapplied,preprint,superscriptaddress]{revtex4-2}

\usepackage{graphicx}
\usepackage{dcolumn}
\usepackage{bm}
\usepackage{xcolor}

\begin{document}

\title{Strong acoustic phonon suppression leads to ultralow  thermal conductivity and enhanced thermoelectric performance in BaCuGdTe$_3$}

\author{Jyoti Duhan}
\affiliation{Department of Physics, Indian Institute of Technology Kanpur, UP 208016, India}
\author{Chris Wolverton}
\affiliation{Department of Materials Science and Engineering, Northwestern University, Evanston, Illinois 60208, United States}
\author{Koushik Pal \footnote[1]{koushik@iitk.ac.in}}
\affiliation{Department of Physics, Indian Institute of Technology Kanpur, UP 208016, India}

\date{\today}

\begin{abstract}
Excitations and scatterings among the quantized lattice vibrations, i.e., phonons, govern the lattice thermal conductivity ($\kappa_l$) in crystalline solids. Therefore, effective modulation of $\kappa_l$ can be achieved through selective manipulation of phonon modes that strongly participate in the heat transport mechanisms. Here, combining accurate first-principles density functional theory calculations and Boltzmann transport theory, we report a layered quaternary chalcogenide semiconductor, BaCuGdTe$_3$, which exhibits unusually low $\kappa_l$ ($\sim$ 0.14 W/m·K at room temperature) despite its ordered crystalline structure. Our analysis reveals that the ultralow $\kappa_l$ arises mainly from a strong suppression of acoustic phonon modes induced by local distortion, shear vibrations among the layers, and large acoustic-optical avoided-crossing between phonons, which collectively enhances the phonon-scattering rates. Further calculations of the electrical transport properties with explicit consideration of electron-phonon interactions reveal a high thermoelectric figure of merit exceeding unity for this compound at moderate temperature (400-700 K) and carrier concentration \((1–5 \times 10^{19} \ \text{cm}^{-3})\) ranges. Our theoretical predictions warrant experimental investigations of the intriguing phonon dynamics, thermal transport mechanisms, and thermoelectric properties in this compound. Moreover, insights from our analysis can be used to design and engineer compounds with ultralow $\kappa_l$. 

\end{abstract}

\maketitle

\section{Introduction}

With the increasing focus on utilizing green energy sources to mitigate environmental damage, thermoelectric materials offer potential sustainable solutions for their ability to convert heat into electrical energy \cite{bell2008cooling,snyder2008complex}. Thermoelectric materials have diverse applications, including power generation utilizing the wasted heat generated by fossil fuel combustion, cooling, heat pumping, and thermal sensors \cite{yang2018high, shi2014novel, tan2019thermoelectric, zhao2014review, riffat2006performance}. The efficiency of a thermoelectric material is characterized by the dimensionless figure of merit, denoted by \(ZT\), expressed as $ZT = S^2\sigma T/(\kappa_e + \kappa_l)$, where \(S\) represents the Seebeck coefficient, \(\sigma\) denotes the electrical conductivity, \(T\) is the temperature,  \(\kappa_e\)\ is the electronic thermal conductivity, and \(\kappa_l\)\ is the lattice thermal conductivity. The Wiedemann-Franz law relates the electronic thermal conductivity to electrical conductivity through the relation $\kappa_e\ = L \sigma T$, where \(L\) is the Lorenz number. Thermoelectric materials are usually small band gap semiconductors ($\sim$ 0.1 - 1 eV) which should possess simultaneously high electrical transport properties ($S$, $\sigma$) and low thermal conductivity ($\kappa$ = $\kappa_e$ + $\kappa_l$). Since $\kappa_e$ and $\sigma$ are directly related, it is often advantageous to minimize the lattice thermal conductivity while maintaining high electrical conductivity. To enhance the efficiency of thermoelectrics, G. A. Slack proposed the concept of phonon-glass electron-crystal (PGEC) \cite{rowe2018crc} to optimize the figure of merit in which phonons responsible for thermal conductivity should experience the material as glass, with a high degree of phonon scattering to lower thermal conductivity. Meanwhile, electrons should experience the material as a crystal, with minimal scattering to maintain high electrical conductivity.

PGEC behavior was initially observed in cage-like compounds such as skutterudites \cite{shi2011multiple} and clathrates \cite{takabatake2014phonon}, which are known for their low $\kappa_l$. Hence in addition to band engineering, which enhances electrical transport properties, various strategies have been explored to modulate $\kappa_l$, including nano-crystallization \cite{liu2017new}, introducing crystal defects \cite{chen2017vacancy}, inducing structural disorders \cite{li2018liquid}, and incorporating rattling guest atoms \cite{jana2017intrinsic}. However, these approaches may compromise electronic transport properties. Hence, finding crystalline semiconductors with intrinsically low $\kappa_l$ is a good choice for designing high-performance thermoelectrics. A wide range of crystalline chalcogenides, including SnSe \cite{zhao2014review, zhao2016ultrahigh, chang20183d}, GeTe \cite{perumal2019realization, samanta2019realization, jiang2022high}, CsBi$_4$Te$_6$ \cite{chung2000csbi4te6}, AgSbTe$_2$ \cite{roychowdhury2021enhanced}, Yb$_{14}$MnSb$_{11}$ \cite{brown2006yb14mnsb11, perez2021discovery}, Mg$_3$(Si, Bi)$_2$ \cite{ding2021soft}, CsAg$_5$Te$_3$ \cite{lin2016concerted} and La$_{3-x}$Te$_4$ \cite{may2009influence}, have been extensively studied for their ultralow $\kappa_l$ and large thermoelectric figure of merit.

Recently, a family of quaternary chalcogenides, denoted by AMM$^{\prime}$Q$_3$ (A = alkali, alkaline earth, post-transition metals; M/M$^{\prime}$ = transition metals, lanthanides, actinides; Q = chalcogens) has gained a lot of attention for their interesting structure-property relationship \cite{prakash2015syntheses, berseneva2023advances, pal2019unraveling}. Many of the semiconductors in this family have been shown to exhibit low thermal conductivity \cite{pal2021accelerated, pal2019intrinsically, pal2022scale, hao2019design, shahid2023synthesis, yu2024substitution, laing2022acuzrq3, rohj2024ultralow}. The reduction in $\kappa_l$ in this family is primarily due to  (i) the presence of the rattler cations (A$^{m+}$) at the A site, which generates localized phonon modes that introduce additional phonon-scattering channels and (ii) the large anharmonicity, which increases phonon-phonon scattering rates. These combined effects make AMM$^{\prime}$Q$_3$ chalcogenides highly effective at reducing heat transport through the lattice while maintaining electrical conductivity due to strong covalent bonding present within the [MM$^{\prime}$Q$_3$]$^{m-}$ layers \cite{pal2019high}. These promising characteristics of AMM$^{\prime}$Q$_3$ chalcogenides motivated us to search for high-performance thermoelectric materials in this family. One recent experimental study reports a relatively high \(ZT\) of 0.8 in BaCuScTe$_3$ \cite{yu2024substitution} at 773K. In this study, we choose BaCuGdTe$_3$, a close cousin of BaCuScTe$_3$ for detailed calculations and analysis.  This choice stems from a thorough comparative analysis of  their crystal  structures, chemical bonding, thermodynamic and phonon related properties. Our analysis (see Supplemental Material \cite{supplement} Section II) reveals that BaCuGdTe$_3$ possesses characteristics conducive to host lattice thermal conductivity lower than that of BaCuScTe$_3$.


\section{Structural Properties}

BaCuGdTe$_3$ exhibits a base-centered orthorhombic structure (space group Cmcm, \#63) \cite{prakash2015syntheses}. The Ba, Cu, and Gd occupy the 4c, 4c, and 4a Wyckoff sites, respectively, and the chalcogen atom Te occupy the 4c and 8f sites. The BaCuGdTe$_3$ has a layered structure with parallel [CuGdTe$_3$]$^{2-}$ slabs separated from each other by Ba$^{2+}$ atoms. Gd$^{3+}$ is bonded to six Te$^{2-}$ atoms to form GdTe$_6$ octahedra sharing corners and edges with equivalent polyhedra, while Cu$^{1+}$ is bonded to four Te$^{2-}$ atoms to form CuTe$_4$ tetrahedra sharing corners with equivalent tetrahedra and edges with GdTe$_6$ octahedra. GdTe$_6$ octahedra have two bonds of 3.07 Å and four bonds of 3.12 Å, while CuTe$_4$ tetrahedra have bonds of 2.67 Å and 2.71 Å. The tetrahedral Te–Cu–Te and octahedral Te–Gd–Te angles deviate slightly from the ideal tetrahedral and octahedral angles making these polyhedra distorted. Structural distortions are shown to be extremely effective in reducing lattice thermal conductivity in crystalline materials such as $\beta$-Cu$_2$Se \cite{kim2015ultralow}, AgBiS$_2$ \cite{rathore2019origin}, and CuFeS$_2$ \cite{tippireddy2022local}. Therefore, BaCuGdTe$_3$ is also expected to exhibit low $\kappa_l$ due to its local structural distortion.

\begin{figure*}[t]
\centering
\includegraphics[width=0.99\textwidth]{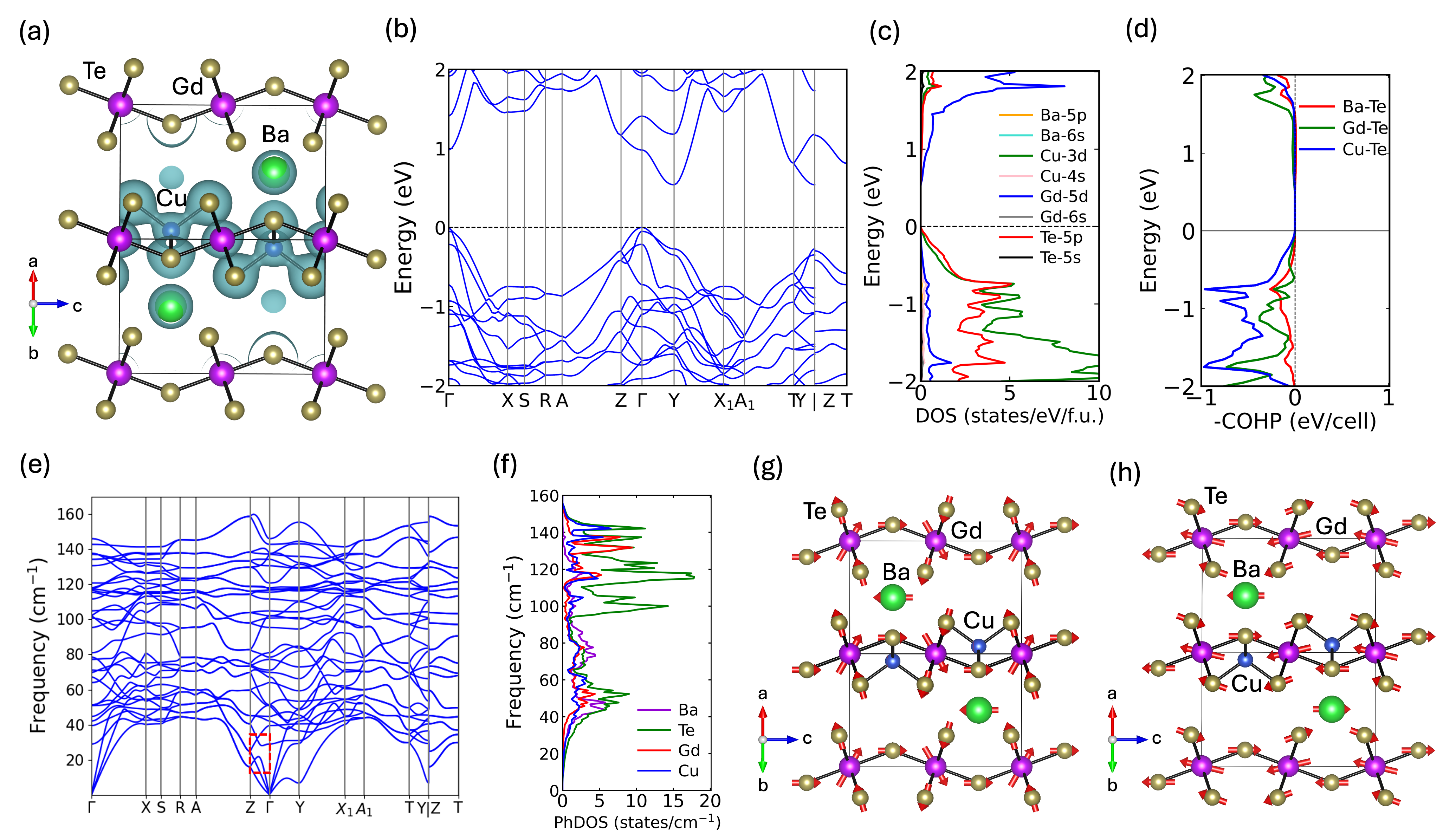}
\caption{(a) Total charge density distribution visualized at an isosurface value of $0.04 \, \text{e}/bohr^3$, (b) electronic band structure, (c) partial density of states (PDOS), (d) crystal orbital Hamilton population (COHP) analysis,  (e) phonon dispersion (the red-dashed square highlights the avoided crossing between the LA mode and the lowest-frequency optical phonon mode), (f) phonon density of states (PhDOS) of BaCuGdTe$_3$. (g) Visualizations of the eigenvector of the lowest energy optical phonon mode at $\Gamma$ point (29 $cm^{-1}$), and (h) the lowest energy acoustic phonon mode at the Y point (7 $cm^{-1}$). Arrows on the atoms indicate the direction of vibration.}
\end{figure*}

To understand the structure and properties of this compound, we performed detailed first-principles theoretical calculations based on density functional theory (DFT). The details of the calculations are provided in the Section I of Supplemental Material \cite{supplement}. To understand the chemical bonding in this compound, we visualize the total charge density in Fig. 1(a) which shows a strong overlap of charge clouds between the atoms in the [CuGdTe$_3$]$^{2-}$ layers indicating covalent interactions. While the charge densities around the cations Ba$^{2+}$ remain isolated and almost spherical, highlighting its ionic bonding nature that interacts with the covalently bonded layers through electrostatic interactions. Calculated electronic structure [Fig. 1(b)] reveals that this compound possesses an indirect band gap of 0.54 eV, where the valence band maximum (VBM) is located at the $\Gamma$ point while the conduction band minimum (CBM) is at the Y point in the Brillouin zone (BZ). A slight energy split of 0.05 eV is also observed between the VBM and the secondary VBM at $\Gamma$ point in the BZ. 
This small energy split indicates that the secondary valence band might influence the electronic transport properties through band convergence, which can enhance the Seebeck coefficient by increasing the density of states effective mass (see Section III of Supplemental Material \cite{supplement} for detailed explanantion). The projected density of states (PDOS) in Fig. 1(c) shows that the valence band is primarily composed of Cu-3d and Te-5p states, while the conduction band has a major contribution from Gd-5d states followed by a contribution from Te-5p states. To further understand the interactions between the atoms, we performed the crystal orbital Hamilton population (COHP) analysis shown in Fig. 1(d). In the COHP plot, the positive and negative values on the x-axis represent bonding and antibonding states, respectively following the original convention \cite{dronskowski1993crystal}. The analysis reveals strong antibonding interactions below the Fermi level between Cu and Te atoms, and between Gd and Te atoms, while the interaction between Ba and Te is mildly antibonding. The presence of antibonding states at the top of the valence bands is strongly related to ultralow lattice thermal conductivity in solids \cite{ubaid2024antibonding, yuan2023lattice, das2023strong, yuan2022antibonding}. The presence of different bonding interactions and strengths within a compound forms a chemical bonding hierarchy, which can also contribute to improved thermoelectric performance \cite{pal2018bonding, eickmeier2022exploring, biswas2012high}.

\section{Phonon Transport}

Phonon dispersion and atom-projected phonon density of states for BaCuGdTe$_3$ are shown in Figs. 1(e) and 1(f), respectively. We notice several interesting features in the phonon dispersion. The cut-off frequencies $(\omega_{cut})$ for the acoustic phonon branches are very small in all directions in the BZ. For example, along $\Gamma$-X, $\omega_{cut}$ for the longitudinal acoustic (LA) branch is 48 $cm^{-1}$ which is highest among all directions. The lowest $\omega_{cut}$ (= 9 $cm^{-1}$) is seen for the transverse acoustic (TA) branch along the $\Gamma$-Y direction, which corresponds to the layer direction in the crystal structure. Along $\Gamma$-Z, the maximum $\omega_{cut}$ is 25 $cm^{-1}$ which is found at the Z point for one of the TA branches. These low $\omega_{cut}$ signify very small speeds of sound that should give rise to very low $\kappa_l$. Along the $\Gamma$-Z direction, we found a large avoided crossing between LA and lowest frequency optical phonon branches, that suppresses the $\omega_{cut}$ for the LA mode to 16 $cm^{-1}$ and subsequently reduces the sound speed. Interestingly, we see that one of the TA phonon branches along the $\Gamma$-Y direction is strongly suppressed whose $\omega_{cut}$ becomes very low (9 $cm^{-1}$). This phonon branch dips further in energy at the Y point to 7 $cm^{-1}$ and then rises again along the Y-$X_1$ direction. This strong suppression is caused by the layered structure and local distortion present in the covalently bonded layers. This damped phonon branch is expected to reduce the phonon group velocity and enhance the phonon scattering rates, inducing ultralow $\kappa_l$ \cite{wang2015anisotropic,li2020ultralow}. Additionally, we notice that many nearly dispersionless phonon branches are localized near 50 $cm^{-1}$, which arise from  the large amplitude vibrations of weakly bound Ba atoms, akin to rattling modes. These low-energy phonon modes are also expected to give rise to strong phonon-scattering, reducing the $\kappa_l$ in BaCuGdTe$_3$ \cite{jana2016origin}. Further, we have analysed atom-resolved phonon dispersion, shown in Supplemental Material \cite{supplement} Fig. S4(a), to understand the contributions of various atoms along different phonon branches. The atom-projected phonon density of states (PhDOS) and atom-resolved phonon dispersion reveal that the contribution of Ba atoms around 50 $cm^{-1}$ overlaps with Te.

The avoided crossing observed along $\Gamma$-Z direction between the highest-frequency acoustic phonon mode (at approximately 22 $cm^{-1}$), is predominantly contributed by Ba and Te atoms. While, the lowest-frequency optical phonon mode (near 28 $cm^{-1}$) along the same direction mainly involves Gd and Te atoms. This interpretation is further supported by the visualization of the corresponding phonon eigenvectors in Supplemental Material \cite{supplement} Figs. S4(b,c), which clearly reflect the atomic displacement contributing to these modes. Further, we present visualizations of the eigenvectors of the lowest energy optical phonon mode at $\Gamma$ point (29 $cm^{-1}$) and the lowest energy acoustic phonon mode at Y point (7 $cm^{-1}$) in Figs. 1(g) and 1(h), respectively. Figure 1(g) shows the vibration of the Ba atoms that is admixed with relatively larger vibrations of Gd and smaller contributions of Te atoms. Figure 1(h) shows the shear-like vibration of atoms in the plane parallel with the covalently bonded layers. Both avoided crossing in phonon branches and shear motion are expected to be effective in scattering heat-carrying phonons, reducing $\kappa_l$ to ultralow value \cite{christensen2008avoided, jana2016origin}.

\begin{figure*}[t]
\centering
\includegraphics[width=0.9\textwidth]{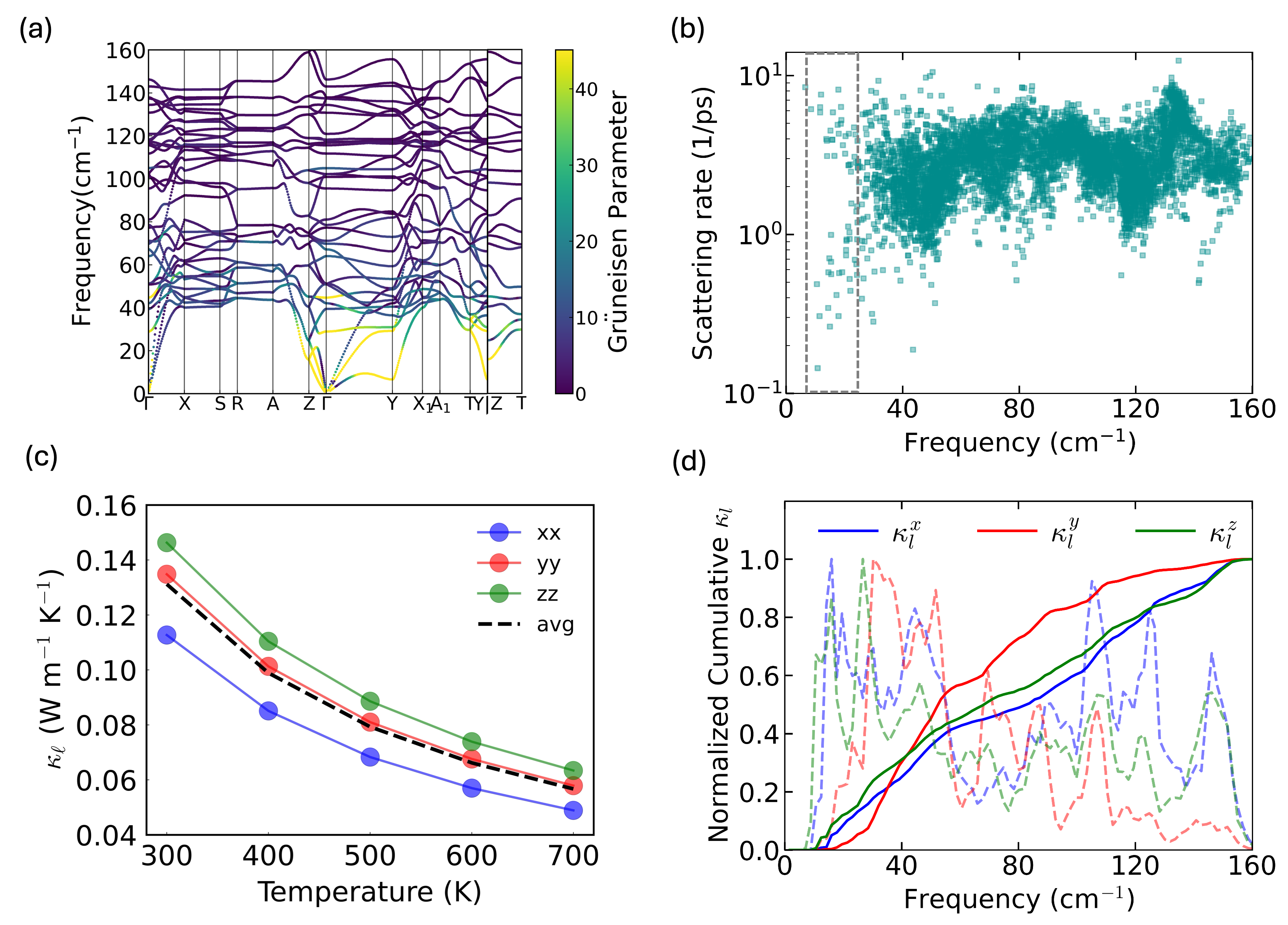}
\caption{(a) Mode resolved Grüneisen parameters ($\gamma_{qv}$), which are color coded on phonon dispersion of BaCuGdTe$_3$, where the color bar represents the magnitude. (b) Three-phonon scattering rates at 300K, (c) temperature-dependent lattice thermal conductivity ($\kappa_l$), and (d) normalized cumulative $\kappa_l$ with respect to the phonon frequency and their derivatives (represented by dotted line) at 300K, along different crystallographic directions for BaCuGdTe$_3$. Large scattering rates which arise from the shear vibrations and avoided-crossing regions  are highlighted by the grey-dashed rectangle.}
\end{figure*}

The lattice thermal conductivity of a crystalline solid is expressed by the relation $ \kappa_l = \frac{1}{3} C_v v^2 \tau $, where \( C_v \) represents the specific heat capacity of the system at constant volume, \( v \) is the phonon group velocity, and \( \tau \) denotes the phonon lifetime. The low $\omega_{cut}$ of the acoustic modes (below 50 $cm^{-1}$) suggests low group velocities for the compound, which support ultralow $\kappa_l$. Next, we calculate the Grüneisen parameter ($\gamma_{qv}$) for the phonon modes of BaGdCuTe$_3$ [Fig. 2(a)], which shows very high values of $\gamma_{qv}$ $(>>1)$ for phonon frequencies up to 60 $cm^{-1}$. For phonon modes where avoided crossing and shear motion appear, the $\gamma_{qv}$  values become extremely large (30-40). Since Grüneisen parameter measures the degree of anharmonicity in a system, the higher $\gamma_{qv}$$^\prime$s lead to stronger phonon-phonon scattering, which lowers the $\kappa_l$ \cite{xiao2024anomalous, chang2018anharmoncity}.

We now calculate the phonon-scattering rates and determine the $\kappa_l$ by solving the iterative solution of the linearized Boltzmann transport equation (see computation details in the Supplemental Material \cite{supplement}). The calculated scattering rates are presented in Fig. 2(b) which shows large values in the entire frequency range, signifying the fact that not only the low energy acoustic and optical phonon modes (up to 60 cm$^{-1}$) but also high energy phonons $(> 60 cm^{-1})$ actively participate in the thermal transport mechanism. Notably, in the low-frequency regime, we notice large scattering rates of phonon modes near the avoided crossing (25 cm$^{-1}$) and near the suppressed phonon modes (7 cm$^{-1}$) due to shear vibration, as highlighted by the grey-dashed rectangle in Fig. 2(b). The mode-resolved scattering rates, as shown in Supplemental Material \cite{supplement} Figs. S5(b,c), indicate that the scattering rates along the $\Gamma$-Z direction (where the avoided crossing occurs) are higher than along any arbitary direction in the Brillouin zone. The calculated $\kappa_l$ is ultralow and  anisotropic as shown in Fig. 2(c), with values of 0.11 W/m·K along the x-axis, 0.13 W/m·K along the y-axis, and 0.15 W/m·K along the z-axis at room temperature. These values are smaller than some of the ultralow $\kappa_l$ quaternary compounds such as AgPbBiSe$_3$ (0.5 W/m.K) \cite{dutta2019bonding}, AgSnSbTe$_3$ (0.47 W/m.K) \cite{sarkar2023chemical} and also lower than SnSe (0.34 W/m.K) \cite{rundle2022layered, zhao2016ultrahigh}. The anisotropy is arising from the layered crystal structure of BaCuGdTe$_3$. As the temperature increases, the $\kappa_l$ decreases due to enhanced phonon-phonon scattering. To understand the mode contribution to $\kappa_l$, we plot energy cumulative $\kappa_l$ and its derivative in Fig. 2(d) along different directions. Along the x and z directions, we can see that both the acoustic and optical phonon modes up to 150 $cm^{-1}$ are contributing to $\kappa_l$, while along the y direction, the frequencies above 30 $cm^{-1}$ are mainly contributing. The derivative of the cumulative $\kappa_l$ indicates that the major contribution to $\kappa_l$ comes from frequencies up to 60 cm$^{-1}$, which is supported by the high values of $\gamma_{qv}$$^\prime$s in this frequency range. The contribution from phonon modes above 60 cm$^{-1}$ is also significant, as revealed by their high scattering rates.

\section{Electrical Transport}

We now turn our attention to calculate and analyze the electrical transport properties namely, electrical conductivity ($\sigma$), Seebeck coefficient ($S$), and power factor ($S^2 \sigma$). We calculated $\sigma$, $S$, and $\kappa_e$ by solving the linearized Boltzmann transport equation using the AMSET code \cite{ganose2021efficient} considering both p-type and n-type carriers. Detailed information about the calculations and convergence tests  can be found in the Supplemental Material \cite{supplement} Fig. S7. We present the results for $\sigma$ [Figs. S8(a,d)], $S$ [Figs. S8(b,e)] and $\kappa_e$ [Fig. S9] in the Supplemental Material \cite{supplement}. BaCuGdTe$_3$ exhibits anisotropic effective masses for electrons compared to the holes. The calculated hole effective masses at the VBM are 0.597 m$_0$ along $\Gamma$-X, 0.717 m$_0$ along $\Gamma$-Y, and 0.763 m$_0$ along $\Gamma$-Z directions. For electrons at the CBM, the effective masses are 0.344 m$_0$ along Y-$\Gamma$, 0.760 m$_0$ along Y-X$_1$, and 0.751 m$_0$ along Y-T directions. As a result, the average effective mass of holes is 0.692 m$_0$, while the average effective mass of electrons is 0.618 m$_0$. Since effective masses are inversely related to carrier mobility, this implies that electron conductivity will be higher than hole conductivity due to the lower effective mass in CBM. These observations are consistent with the results shown in Figs. S10(b,e,h) and S11(b,e,h). Due to the low effective electron mass along the Y-$\Gamma$ direction, the electrical conductivity reaches a high value of 4582 $\mathrm{S/cm}$ at a carrier concentration of \(5 \times 10^{20} \ \text{cm}^{-3}\) at room temperature [Fig. S10(b) in the Supplemental Material \cite{supplement}]. Further, the small energy difference between the VBM and secondary VBM indicates that this will influence  the electronic transport properties through changing the density of states effective mass (\( m_{\text{dos}}^* \)). Our analysis also shows that a reduction in the energy difference between those two valence bands can be achieved through strain engineering which increases the \( m_{\text{dos}}^* \),  giving rise to a  higher S and hence, higher thermoelectric performance (see Supplemental Material Section III \cite{supplement}).

We calculated the electronic transport properties taking a carrier concentration range from \(1 \times 10^{19}\) to \(5 \times 10^{20} \ \text{cm}^{-3}\) and a temperature range of 300\,\text{--}\,700\,\text{K}, which are typical for thermoelectric materials. The direction-averaged Seebeck coefficient at room temperature is \( 253 \, \mu \text{V}/\text{K} \) for p-type doping [Fig. S8(b) in the Supplemental Material \cite{supplement}] and \( -236 \, \mu \text{V}/\text{K} \) for n-type doping [Fig. S8(e) in the Supplemental Material \cite{supplement}] at a concentration of \(1 \times 10^{19} \ \text{cm}^{-3}\). These values exceed room temperature $S$ of commonly known thermoelectric materials such as SnSe, which has \(S = -180 \, \mu \text{V}/\text{K}\) (n-type) and \(S = 210 \, \mu \text{V}/\text{K}\) (p-type) at a doping concentration of \(1.2 \times 10^{19} \, \text{cm}^{-3}\) \cite{chang20183d}, and PbTe, which has \(S = 178 \, \mu \text{V}/\text{K}\) at a concentration of \(3.6 \times 10^{19} \, \text{cm}^{-3}\) \cite{kim2012effect}. We notice that $S$ exhibits non-monotonic behavior for both p-type [Fig. S8(b) in the Supplemental Material \cite{supplement}] and n-type  [Fig. S8(e) in the Supplemental Material \cite{supplement}] carriers at low concentrations and higher temperatures due to the bipolar conduction effect \cite{shahi2018bipolar}. This effect occurs when the temperature is sufficiently high to excite charge carriers across the bandgap, leading to a reduction in the Seebeck coefficient.

The electrical conductivity ($\sigma$) decreases with temperature but increases with doping concentration [Figs. S8(a) and S8(d) in the Supplemental Material \cite{supplement}]. The direction averaged $\sigma$ reaches a peak value of 2318 $\mathrm{S/cm}$ for n-type and 1156 $\mathrm{S/cm}$ for p-type at room temperature, while at 700K, it drops to 1103 $\mathrm{S/cm}$ for n-type and 529 $\mathrm{S/cm}$ for p-type doping concentration of \(5 \times 10^{20} \ \text{cm}^{-3}\). The Seebeck coefficient generally shows an opposite dependence on temperature and carrier concentration compared to electrical conductivity. Consequently, to optimize the power factor, defined as \( S^2 \sigma \), we must achieve a balance between $S$ and $\sigma$ as a function of carrier concentration and temperature so that their product $S^2\sigma$ becomes maximum. The calculated electrical thermal conductivity (\(\kappa_e\)) increases with doping concentrations. As shown in Fig. S9 in the Supplemental Material \cite{supplement}, at room temperature, the electronic thermal conductivity is 0.035 W/m·K for the n-type and 0.028 W/m·K for the p-type doping concentration of \(1 \times 10^{19} \ \text{cm}^{-3}\). The value of $\kappa_e$ for a particular doping concentration remains nearly constant across the temperature range. 

\section{Figure of Merit}

\begin{figure*}[t]
\centering
\includegraphics[width=0.9\textwidth]{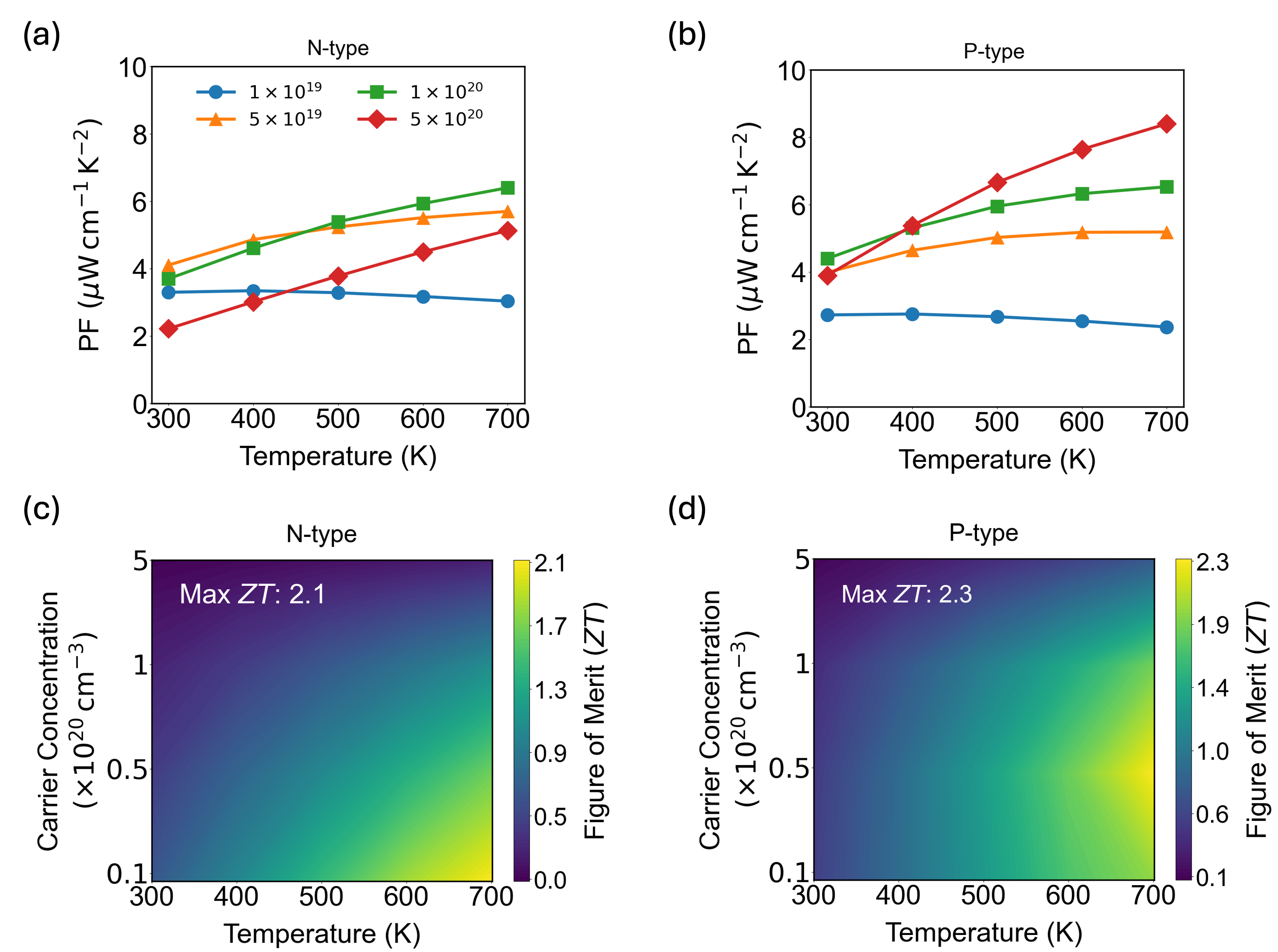}
\caption{Calculated power factor (a,b) and thermoelectric figure of merit (c,d) as a function of temperature and carrier concentration for both n- and p-doped $\text{BaCuGdTe}_3$.}
\end{figure*}

By combining the electrical and thermal transport properties, we now predict the power factor and dimensionless figure of merit $ZT$ as a function of temperature and carrier concentration. The power factor (PF) for the n-type and p-type BaCuGdTe\(_3\) are shown in Figs. 3(a) and 3(b), respectively. For n-type doping, the maximum power factor is \(6.4 \, \mu \text{W} \, \text{cm}^{-1} \, \text{K}^{-2}\), achieved at 700 K and a carrier concentration of \(1 \times 10^{20} \, \text{cm}^{-3}\). For p-type doping, the maximum power factor is \(8.4 \, \mu \text{W} \, \text{cm}^{-1} \, \text{K}^{-2}\), observed at 700 K and a carrier concentration of \(5 \times 10^{20} \, \text{cm}^{-3}\). These values are comparable to those reported for SnSe (\(5.65 \, \mu \text{W} \, \text{cm}^{-1} \, \text{K}^{-2}\) for p-type and \(7.66 \, \mu \text{W} \, \text{cm}^{-1} \, \text{K}^{-2}\) for n-type at a carrier concentration of \(1.2 \times 10^{19} \, \text{cm}^{-3}\) at 300 K) \cite{gong2018extremely,chang20183d}, and CuPbSbSe$_3$ (\(3.2 \, \mu \text{W} \, \text{cm}^{-1} \, \text{K}^{-2}\) for n-type and \(14.2 \, \mu \text{W} \, \text{cm}^{-1} \, \text{K}^{-2}\) for p-type at a carrier concentration of \(1 \times 10^{20} \, \text{cm}^{-3}\) and 600 K) \cite{faghaninia2017computational}. Figures S10(c,f,j) and S11(c,f,j) illustrate that the PFs in the x and z directions are significantly greater than in the y direction due to layered structure and stronger bonding within the [CuGdTe$_3$]$^{2-}$ layers, resulting in enhanced electronic transport along the in-plane direction. The calculated dimensionless figure of merit $ZT$ is shown in Figs. 3(c) and 3(d) show that for p-doped $\text{BaCuGdTe}_3$, a maximum average $ZT$ of 2.3 is predicted at a carrier concentration of $5 \times 10^{19} \ \text{cm}^{-3}$ and a temperature of 700 K. Meanwhile, for n-doped $\text{BaCuGdTe}_3$, a maximum average $ZT$ of 2.1 is obtained at a carrier concentration of $1 \times 10^{19} \ \text{cm}^{-3}$ and a temperature of 700 K. Figure S12 in Supplemental Material \cite{supplement}  presents the maximum $ZT$ values along different directions as a function of carrier concentration and temperature. These high figure of merit values signify that BaCuGdTe\(_3\) is a promising thermoelectric material that should be investigated in detail experimentally. An experimental study \cite{gunatilleke2023structure} on the same composition reports temperature-dependent (in the 40-300 K range) resistivity ($\rho$) and Seebeck coefficient measurements and found small values of $S$ and $\rho$, consistent with metallic behavior. However, our  electronic structure calculations, along with a previous study \cite{eickmeier2022exploring}, describe the material as a semiconductor with band gaps of 0.54 eV and 0.56 eV, respectively. Even when accounting for the unscreened Gd 4f-orbital in the pseudopotential, which lowers the band gap, the material remains semiconducting [Fig. S13 in the Supplemental Material  \cite{supplement}]. The observed metallic behavior in the experiment \cite{gunatilleke2023structure} can arise from the unwarranted presence of defects in the sample. Therefore, careful synthesis methods and analysis of defects should be undertaken to realize the high thermoelectric performance of this compound. 

\section{Conclusions}

In conclusion, using accurate first-principles calculations and linearized Boltzmann transport theory for phonons, we predict an ultralow lattice thermal conductivity in  crystalline BaCuGdTe$_3$. Our analysis reveals that local structural distortions can strongly suppress the acoustic phonon branches, which can significantly enhance phonon scattering and reduce the phonon group velocities. Additionally, the presence of low-energy shear vibrational modes  and  localized  phonon branches due to the vibrations of weakly bound cations can further amplify the phonon scattering channels, reducing the lattice thermal conductivity to an ultralow value.  Further, calculations of electrical transport properties show a relatively high Seebeck coefficient and electrical conductivity in this compound. While structural distortions and weakly bound Ba facilitate a poor thermal transport, the presence of strong covalent bonding along the [CuGdTe$_3$]$^{2-}$ layer gives rise to enhanced electrical transport properties. As a result, BaCuGdTe$_3$ demonstrates a favorable combination of ultralow thermal conductivity and a relatively high power factor, making it a potential material for thermoelectric applications with high \(ZT\) values at moderate carrier concentrations and temperatures. Our analysis further reveals that strain engineering of electronic bands can be a viable tool to control the energy splitting between the two topmost valence bands at $\Gamma$ point, giving rise to enhanced density of states effective mass, Seebeck coefficient and potentially high thermoelectric performance.
 
\section{Acknowledgements}

J.D. is grateful for the financial support provided by IIT Kanpur through the institute fellowship. C.W. acknowledges support from the Department of Energy, Office of Science, Basic Energy Sciences under grant DE-SC0014520. K.P. acknowledges financial support from an Initiation Grant from the IIT Kanpur. We acknowledge computational resources provided by the (a) HPC2013 and Param Sanganak computing facilities provided by IIT Kanpur. 

\section{Data Availability}

{The data that support the findings of this study are given in the figures and table, which are available from the authors upon reasonable request.}

\bibliographystyle{unsrt}
\bibliography{ref}

\begin{thebibliography}{10}

\bibitem{bell2008cooling}
Lon~E Bell.
\newblock Cooling, heating, generating power, and recovering waste heat with thermoelectric systems.
\newblock {\em Science}, 321(5895):1457--1461, 2008.

\bibitem{snyder2008complex}
G~Jeffrey Snyder and Eric~S Toberer.
\newblock Complex thermoelectric materials.
\newblock {\em Nature Materials}, 7(2):105--114, 2008.

\bibitem{yang2018high}
Lei Yang, Zhi-Gang Chen, Matthew~S Dargusch, and Jin Zou.
\newblock High performance thermoelectric materials: progress and their applications.
\newblock {\em Advanced Energy Materials}, 8(6):1701797, 2018.

\bibitem{shi2014novel}
Yongming Shi, Yao Wang, Yuan Deng, Hongli Gao, Zhen Lin, Wei Zhu, and Huihong Ye.
\newblock A novel self-powered wireless temperature sensor based on thermoelectric generators.
\newblock {\em Energy Conversion and Management}, 80:110--116, 2014.

\bibitem{tan2019thermoelectric}
Gangjian Tan, Michihiro Ohta, and Mercouri~G Kanatzidis.
\newblock Thermoelectric power generation: from new materials to devices.
\newblock {\em Philosophical Transactions of the Royal Society A}, 377(2152):20180450, 2019.

\bibitem{zhao2014review}
Dongliang Zhao and Gang Tan.
\newblock A review of thermoelectric cooling: Materials, modeling and applications.
\newblock {\em Applied Thermal Engineering}, 66(1-2):15--24, 2014.

\bibitem{riffat2006performance}
SB~Riffat, Xiaoli Ma, and Robin Wilson.
\newblock Performance simulation and experimental testing of a novel thermoelectric heat pump system.
\newblock {\em Applied Thermal Engineering}, 26(5-6):494--501, 2006.

\bibitem{rowe2018crc}
David~Michael Rowe.
\newblock {\em CRC handbook of thermoelectrics}.
\newblock CRC press, 2018.

\bibitem{shi2011multiple}
Xun Shi, Jiong Yang, James~R Salvador, Miaofang Chi, Jung~Y Cho, Hsin Wang, Shengqiang Bai, Jihui Yang, Wenqing Zhang, and Lidong Chen.
\newblock Multiple-filled skutterudites: {High} thermoelectric figure of merit through separately optimizing electrical and thermal transports.
\newblock {\em Journal of the American Chemical Society}, 133(20):7837--7846, 2011.

\bibitem{takabatake2014phonon}
Toshiro Takabatake, Koichiro Suekuni, Tsuneyoshi Nakayama, and Eiji Kaneshita.
\newblock Phonon-glass electron-crystal thermoelectric clathrates: Experiments and theory.
\newblock {\em Reviews of Modern Physics}, 86(2):669--716, 2014.

\bibitem{liu2017new}
Weishu Liu, Jizhen Hu, Shuangmeng Zhang, Manjiao Deng, Cheng-Gong Han, and Yong Liu.
\newblock New trends, strategies and opportunities in thermoelectric materials: a perspective.
\newblock {\em Materials Today Physics}, 1:50--60, 2017.

\bibitem{chen2017vacancy}
Zhiwei Chen, Binghui Ge, Wen Li, Siqi Lin, Jiawen Shen, Yunjie Chang, Riley Hanus, G~Jeffrey Snyder, and Yanzhong Pei.
\newblock Vacancy-induced dislocations within grains for high-performance {PbSe} thermoelectrics.
\newblock {\em Nature Communications}, 8(1):13828, 2017.

\bibitem{li2018liquid}
B~Li, H~Wang, Y~Kawakita, Qijing Zhang, M~Feygenson, HL~Yu, D~Wu, K~Ohara, T~Kikuchi, K~Shibata, et~al.
\newblock Liquid-like thermal conduction in intercalated layered crystalline solids.
\newblock {\em Nature Materials}, 17(3):226--230, 2018.

\bibitem{jana2017intrinsic}
Manoj~K Jana, Koushik Pal, Avinash Warankar, Pankaj Mandal, Umesh~V Waghmare, and Kanishka Biswas.
\newblock Intrinsic rattler-induced low thermal conductivity in {Zintl} type {TlInTe$_2$}.
\newblock {\em Journal of the American Chemical Society}, 139(12):4350--4353, 2017.

\bibitem{zhao2016ultrahigh}
Li-Dong Zhao, Gangjian Tan, Shiqiang Hao, Jiaqing He, Yanling Pei, Hang Chi, Heng Wang, Shengkai Gong, Huibin Xu, Vinayak~P Dravid, et~al.
\newblock Ultrahigh power factor and thermoelectric performance in hole-doped single-crystal {SnSe}.
\newblock {\em Science}, 351(6269):141--144, 2016.

\bibitem{chang20183d}
Cheng Chang, Minghui Wu, Dongsheng He, Yanling Pei, Chao-Feng Wu, Xuefeng Wu, Hulei Yu, Fangyuan Zhu, Kedong Wang, Yue Chen, et~al.
\newblock 3d charge and 2d phonon transports leading to high out-of-plane {$ZT$} in n-type {SnSe} crystals.
\newblock {\em Science}, 360(6390):778--783, 2018.

\bibitem{perumal2019realization}
Suresh Perumal, Manisha Samanta, Tanmoy Ghosh, U~Sandhya Shenoy, Anil~K Bohra, Shovit Bhattacharya, Ajay Singh, Umesh~V Waghmare, and Kanishka Biswas.
\newblock Realization of high thermoelectric figure of merit in {GeTe} by complementary {Co}-doping of {Bi} and {In}.
\newblock {\em Joule}, 3(10):2565--2580, 2019.

\bibitem{samanta2019realization}
Manisha Samanta, Tanmoy Ghosh, Raagya Arora, Umesh~V Waghmare, and Kanishka Biswas.
\newblock Realization of both n-and p-type {GeTe} thermoelectrics: electronic structure modulation by {AgBiSe$_2$} alloying.
\newblock {\em Journal of the American Chemical Society}, 141(49):19505--19512, 2019.

\bibitem{jiang2022high}
Binbin Jiang, Wu~Wang, Shixuan Liu, Yan Wang, Chaofan Wang, Yani Chen, Lin Xie, Mingyuan Huang, and Jiaqing He.
\newblock High figure-of-merit and power generation in high-entropy {GeTe}-based thermoelectrics.
\newblock {\em Science}, 377(6602):208--213, 2022.

\bibitem{chung2000csbi4te6}
Duck-Young Chung, Tim Hogan, Paul Brazis, Melissa Rocci-Lane, Carl Kannewurf, Marina Bastea, Ctirad Uher, and Mercouri~G Kanatzidis.
\newblock {CsBi$_4$Te$_6$}: A high-performance thermoelectric material for low-temperature applications.
\newblock {\em Science}, 287(5455):1024--1027, 2000.

\bibitem{roychowdhury2021enhanced}
Subhajit Roychowdhury, Tanmoy Ghosh, Raagya Arora, Manisha Samanta, Lin Xie, Niraj~Kumar Singh, Ajay Soni, Jiaqing He, Umesh~V Waghmare, and Kanishka Biswas.
\newblock Enhanced atomic ordering leads to high thermoelectric performance in {AgSbTe$_2$}.
\newblock {\em Science}, 371(6530):722--727, 2021.

\bibitem{brown2006yb14mnsb11}
Shawna~R Brown, Susan~M Kauzlarich, Franck Gascoin, and G~Jeffrey Snyder.
\newblock {Yb$_{14}$MnSb$_{11}$}: New high efficiency thermoelectric material for power generation.
\newblock {\em Chemistry of Materials}, 18(7):1873--1877, 2006.

\bibitem{perez2021discovery}
Christopher~J Perez, Maxwell Wood, Francesco Ricci, Guodong Yu, Trinh Vo, Sabah~K Bux, Geoffroy Hautier, Gian-Marco Rignanese, G~Jeffrey Snyder, and Susan~M Kauzlarich.
\newblock Discovery of multivalley fermi surface responsible for the high thermoelectric performance in {Yb$_{14}$MnSb$_{11}$} and {Yb$_{14}$MgSb$_{11}$}.
\newblock {\em Science Advances}, 7(4):eabe9439, 2021.

\bibitem{ding2021soft}
Jingxuan Ding, Tyson Lanigan-Atkins, Mario Calder{\'o}n-Cueva, Arnab Banerjee, Douglas~L Abernathy, Ayman Said, Alexandra Zevalkink, and Olivier Delaire.
\newblock Soft anharmonic phonons and ultralow thermal conductivity in {Mg$_3$(Sb, Bi)$_2$} thermoelectrics.
\newblock {\em Science Advances}, 7(21):eabg1449, 2021.

\bibitem{lin2016concerted}
Hua Lin, Gangjian Tan, Jin-Ni Shen, Shiqiang Hao, Li-Ming Wu, Nicholas Calta, Christos Malliakas, Si~Wang, Ctirad Uher, Christopher Wolverton, et~al.
\newblock Concerted rattling in {CsAg$_5$Te$_3$} leading to ultralow thermal conductivity and high thermoelectric performance.
\newblock {\em Angewandte Chemie International Edition}, 55(38):11431--11436, 2016.

\bibitem{may2009influence}
Andrew~F May, David~J Singh, and G~Jeffrey Snyder.
\newblock Influence of band structure on the large thermoelectric performance of lanthanum telluride.
\newblock {\em Physical Review B—Condensed Matter and Materials Physics}, 79(15):153101, 2009.

\bibitem{prakash2015syntheses}
Jai Prakash, Adel Mesbah, Jessica~C Beard, and James~A Ibers.
\newblock Syntheses and crystal structures of {BaAgTbS$_3$}, {BaCuGdTe$_3$}, {BaCuTbTe$_3$}, {BaAgTbTe$_3$}, and {CsAgUTe$_3$}.
\newblock {\em Zeitschrift f{\"u}r anorganische und allgemeine Chemie}, 641(7):1253--1257, 2015.

\bibitem{berseneva2023advances}
Anna~A Berseneva and Hans-Conrad zur Loye.
\newblock Advances in chalcogenide crystal growth: Flux and solution syntheses, and approaches for postsynthetic modifications.
\newblock {\em Crystal Growth \& Design}, 23(8):5368--5383, 2023.

\bibitem{pal2019unraveling}
Koushik Pal, Xia Hua, Yi~Xia, and Christopher Wolverton.
\newblock Unraveling the structure-valence-property relationships in {AMM$^{\prime}$Q$_3$} chalcogenides with promising thermoelectric performance.
\newblock {\em ACS Applied Energy Materials}, 3(3):2110--2119, 2019.

\bibitem{pal2021accelerated}
Koushik Pal, Yi~Xia, Jiahong Shen, Jiangang He, Yubo Luo, Mercouri~G Kanatzidis, and Chris Wolverton.
\newblock Accelerated discovery of a large family of quaternary chalcogenides with very low lattice thermal conductivity.
\newblock {\em npj Computational Materials}, 7(1):82, 2021.

\bibitem{pal2019intrinsically}
Koushik Pal, Yi~Xia, Jiangang He, and Christopher Wolverton.
\newblock Intrinsically low lattice thermal conductivity derived from rattler cations in an {AMM$^{\prime}$Q$_3$} family of chalcogenides.
\newblock {\em Chemistry of Materials}, 31(21):8734--8741, 2019.

\bibitem{pal2022scale}
Koushik Pal, Cheol~Woo Park, Yi~Xia, Jiahong Shen, and Chris Wolverton.
\newblock Scale-invariant machine-learning model accelerates the discovery of quaternary chalcogenides with ultralow lattice thermal conductivity.
\newblock {\em npj Computational Materials}, 8(1):48, 2022.

\bibitem{hao2019design}
Shiqiang Hao, Logan Ward, Zhongzhen Luo, Vidvuds Ozolins, Vinayak~P Dravid, Mercouri~G Kanatzidis, and Christopher Wolverton.
\newblock Design strategy for high-performance thermoelectric materials: The prediction of electron-doped {KZrCuSe$_3$}.
\newblock {\em Chemistry of Materials}, 31(8):3018--3024, 2019.

\bibitem{shahid2023synthesis}
Omair Shahid, Sweta Yadav, Debanjan Maity, Melepurath Deepa, Manish~K Niranjan, and Jai Prakash.
\newblock Synthesis, crystal structure, {DFT}, and photovoltaic studies of {BaCeCuS$_3$}.
\newblock {\em New Journal of Chemistry}, 47(11):5378--5389, 2023.

\bibitem{yu2024substitution}
Xiaotong Yu, Zhijun Wang, Pei Cai, Kai Guo, Junyan Lin, Shuankui Li, Juanjuan Xing, Jiye Zhang, Xinxin Yang, and Jing-Tai Zhao.
\newblock The substitution of rare-earth {Gd} in {BaScCuTe$_3$} realizing the band degeneracy and the point-defect scattering toward enhanced thermoelectric performance.
\newblock {\em Inorganic Chemistry}, 2024.

\bibitem{laing2022acuzrq3}
Craig~C Laing, Benjamin~E Weiss, Koushik Pal, Michael~A Quintero, Hongyao Xie, Xiuquan Zhou, Jiahong Shen, Duck~Young Chung, Christopher Wolverton, and Mercouri~G Kanatzidis.
\newblock {ACuZrQ$_3$ (A= Rb, Cs; Q= S, Se, Te)}: direct bandgap semiconductors and metals with ultralow thermal conductivity.
\newblock {\em Chemistry of Materials}, 34(18):8389--8402, 2022.

\bibitem{rohj2024ultralow}
Rohit~Kumar Rohj, Animesh Bhui, Shaili Sett, Arindam Ghosh, Kanishka Biswas, and DD~Sarma.
\newblock Ultralow thermal conductivity approaching the disordered limit in crystalline {TlCuZrSe$_3$}.
\newblock {\em Chemistry of Materials}, 2024.

\bibitem{pal2019high}
Koushik Pal, Yi~Xia, Jiangang He, and C~Wolverton.
\newblock High thermoelectric performance in {BaAgYTe$_3$} via low lattice thermal conductivity induced by bonding heterogeneity.
\newblock {\em Physical Review Materials}, 3(8):085402, 2019.

\bibitem{supplement}
See Supplemental Material for computational details on the calculations of thermal and electrical transport properties, which includes Ref. \cite{kresse1996efficiency, kresse1996efficient, kresse1999ultrasoft, setyawan2010high, blochl1994projector, perdew1996generalized, ganose2021efficient, born1926quantenmechanik, frohlich1954electrons, deringer2011crystal, li2014shengbte, li2012thermal, togo2015first, pailhes2014localization, pal2021accelerated, dronskowski1993crystal, parker2015benefits, baur1974geometry}.

\bibitem{kim2015ultralow}
Hyoungchul Kim, Sedat Ballikaya, Hang Chi, Jae-Pyung Ahn, Kiyong Ahn, Ctirad Uher, and Massoud Kaviany.
\newblock Ultralow thermal conductivity of {$\beta$-Cu$_2$Se} by atomic fluidity and structure distortion.
\newblock {\em Acta Materialia}, 86:247--253, 2015.

\bibitem{rathore2019origin}
Ekashmi Rathore, Rinkle Juneja, Sean~P Culver, Nicolo Minafra, Abhishek~K Singh, Wolfgang~G Zeier, and Kanishka Biswas.
\newblock Origin of ultralow thermal conductivity in n-type cubic bulk {AgBiS$_2$}: soft {Ag} vibrations and local structural distortion induced by the {Bi 6s$^2$} lone pair.
\newblock {\em Chemistry of Materials}, 31(6):2106--2113, 2019.

\bibitem{tippireddy2022local}
Sahil Tippireddy, Feridoon Azough, Animesh Bhui, Philip Chater, Demie Kepaptsoglou, Quentin Ramasse, Robert Freer, Ricardo Grau-Crespo, Kanishka Biswas, Paz Vaqueiro, et~al.
\newblock Local structural distortions and reduced thermal conductivity in {Ge}-substituted chalcopyrite.
\newblock {\em Journal of Materials Chemistry A}, 10(44):23874--23885, 2022.

\bibitem{dronskowski1993crystal}
Richard Dronskowski and Peter~E Bl{\"o}chl.
\newblock Crystal orbital {Hamilton} populations ({COHP}): energy-resolved visualization of chemical bonding in solids based on density-functional calculations.
\newblock {\em The Journal of Physical Chemistry}, 97(33):8617--8624, 1993.

\bibitem{ubaid2024antibonding}
Mohammad Ubaid, Paribesh Acharyya, Suneet~K Maharana, Kanishka Biswas, and Koushik Pal.
\newblock Antibonding valence states induce low lattice thermal conductivity in metal halide semiconductors.
\newblock {\em Applied Physics Reviews}, 11(4), 2024.

\bibitem{yuan2023lattice}
Jiaoyue Yuan, Yubi Chen, and Bolin Liao.
\newblock Lattice dynamics and thermal transport in semiconductors with anti-bonding valence bands.
\newblock {\em Journal of the American Chemical Society}, 145(33):18506--18515, 2023.

\bibitem{das2023strong}
Anustoop Das, Koushik Pal, Paribesh Acharyya, Subarna Das, Krishnendu Maji, and Kanishka Biswas.
\newblock Strong antibonding {I(p)--Cu(d)} states lead to intrinsically low thermal conductivity in {CuBiI$_4$}.
\newblock {\em Journal of the American Chemical Society}, 145(2):1349--1358, 2023.

\bibitem{yuan2022antibonding}
Kunpeng Yuan, Xiaoliang Zhang, Zheng Chang, Dawei Tang, and Ming Hu.
\newblock Antibonding induced anharmonicity leading to ultralow lattice thermal conductivity and extraordinary thermoelectric performance in {CsK$_2$X (X= Sb, Bi)}.
\newblock {\em Journal of Materials Chemistry C}, 10(42):15822--15832, 2022.

\bibitem{pal2018bonding}
Koushik Pal, Jiangang He, and C~Wolverton.
\newblock Bonding hierarchy gives rise to high thermoelectric performance in layered {Zintl} compound {BaAu$_2$P$_4$}.
\newblock {\em Chemistry of Materials}, 30(21):7760--7768, 2018.

\bibitem{eickmeier2022exploring}
Katharina Eickmeier, Ruben Poschkamp, Richard Dronskowski, and Simon Steinberg.
\newblock Exploring the impact of lone pairs on the structural features of {Alkaline-Earth (A) Transition-Metal (M, M$^{\prime}$) Chalcogenides (Q) AMM$^{\prime}$Q$_3$}.
\newblock {\em European Journal of Inorganic Chemistry}, 2022(28):e202200360, 2022.

\bibitem{biswas2012high}
Kanishka Biswas, Jiaqing He, Ivan~D Blum, Chun-I Wu, Timothy~P Hogan, David~N Seidman, Vinayak~P Dravid, and Mercouri~G Kanatzidis.
\newblock High-performance bulk thermoelectrics with all-scale hierarchical architectures.
\newblock {\em Nature}, 489(7416):414--418, 2012.

\bibitem{wang2015anisotropic}
Xinjiang Wang, Ruiqiang Guo, Dongyan Xu, JaeDong Chung, Massoud Kaviany, and Baoling Huang.
\newblock Anisotropic lattice thermal conductivity and suppressed acoustic phonons in {MOF-74} from first principles.
\newblock {\em The Journal of Physical Chemistry C}, 119(46):26000--26008, 2015.

\bibitem{li2020ultralow}
Xiyang Li, Peng-Fei Liu, Enyue Zhao, Zhigang Zhang, Tatiana Guidi, Manh~Duc Le, Maxim Avdeev, Kazutaka Ikeda, Toshiya Otomo, Maiko Kofu, et~al.
\newblock Ultralow thermal conductivity from transverse acoustic phonon suppression in distorted crystalline {$\alpha$-MgAgSb}.
\newblock {\em Nature Communications}, 11(1):942, 2020.

\bibitem{jana2016origin}
Manoj~K Jana, Koushik Pal, Umesh~V Waghmare, and Kanishka Biswas.
\newblock The origin of ultralow thermal conductivity in {InTe}: lone-pair-induced anharmonic rattling.
\newblock {\em Angewandte Chemie International Edition}, 55(27):7792--7796, 2016.

\bibitem{christensen2008avoided}
Mogens Christensen, Asger~B Abrahamsen, Niels~B Christensen, Fanni Juranyi, Niels~H Andersen, Kim Lefmann, Jakob Andreasson, Christian~RH Bahl, and Bo~B Iversen.
\newblock Avoided crossing of rattler modes in thermoelectric materials.
\newblock {\em Nature Materials}, 7(10):811--815, 2008.

\bibitem{xiao2024anomalous}
Feng Xiao, Qing-Yu Xie, Xing Ming, Huashan Li, Junrong Zhang, and Bao-Tian Wang.
\newblock Anomalous lattice thermal conductivity of quasi-one-dimensional palladium thiophosphate {A$_2$PdPS$_4$I (A= K, Rb, Cs)}.
\newblock {\em Physical Review B}, 109(24):245202, 2024.

\bibitem{chang2018anharmoncity}
Cheng Chang and Li-Dong Zhao.
\newblock Anharmoncity and low thermal conductivity in thermoelectrics.
\newblock {\em Materials Today Physics}, 4:50--57, 2018.

\bibitem{dutta2019bonding}
Moinak Dutta, Koushik Pal, Umesh~V Waghmare, and Kanishka Biswas.
\newblock Bonding heterogeneity and lone pair induced anharmonicity resulted in ultralow thermal conductivity and promising thermoelectric properties in n-type {AgPbBiSe$_3$}.
\newblock {\em Chemical Science}, 10(18):4905--4913, 2019.

\bibitem{sarkar2023chemical}
Debattam Sarkar, Kapildeb Dolui, Vaishali Taneja, Abdul Ahad, Moinak Dutta, SO~Manjunatha, Diptikanta Swain, and Kanishka Biswas.
\newblock Chemical bonding tuned lattice anharmonicity leads to a high thermoelectric performance in cubic {AgSnSbTe$_3$}.
\newblock {\em Angewandte Chemie}, 135(40):e202308515, 2023.

\bibitem{rundle2022layered}
Jordan Rundle and Stefano Leoni.
\newblock Layered tin chalcogenides {SnS} and {SnSe}: Lattice thermal conductivity benchmarks and thermoelectric figure of merit.
\newblock {\em The Journal of Physical Chemistry C}, 126(33):14036--14046, 2022.

\bibitem{ganose2021efficient}
Alex~M Ganose, Junsoo Park, Alireza Faghaninia, Rachel Woods-Robinson, Kristin~A Persson, and Anubhav Jain.
\newblock Efficient calculation of carrier scattering rates from first principles.
\newblock {\em Nature Communications}, 12(1):2222, 2021.

\bibitem{kim2012effect}
Hyoungchul Kim and Massoud Kaviany.
\newblock Effect of thermal disorder on high figure of merit in {PbTe}.
\newblock {\em Physical Review B—Condensed Matter and Materials Physics}, 86(4):045213, 2012.

\bibitem{shahi2018bipolar}
PJSD Shahi, DJ~Singh, JP~Sun, LX~Zhao, GF~Chen, YY~Lv, J~Li, J-Q Yan, DG~Mandrus, and J-G Cheng.
\newblock Bipolar conduction as the possible origin of the electronic transition in pentatellurides: metallic vs semiconducting behavior.
\newblock {\em Physical Review X}, 8(2):021055, 2018.

\bibitem{gong2018extremely}
Yaru Gong, Cheng Chang, Wei Wei, Jiang Liu, Wenjie Xiong, Shuang Chai, Di~Li, Jian Zhang, and Guodong Tang.
\newblock Extremely low thermal conductivity and enhanced thermoelectric performance of polycrystalline {SnSe} by {Cu} doping.
\newblock {\em Scripta Materialia}, 147:74--78, 2018.

\bibitem{faghaninia2017computational}
Alireza Faghaninia, Guodong Yu, Umut Aydemir, Max Wood, Wei Chen, Gian-Marco Rignanese, G~Jeffrey Snyder, Geoffroy Hautier, and Anubhav Jain.
\newblock A computational assessment of the electronic, thermoelectric, and defect properties of bournonite {(CuPbSbS$_3$)} and related substitutions.
\newblock {\em Physical Chemistry Chemical Physics}, 19(9):6743--6756, 2017.

\bibitem{gunatilleke2023structure}
Wilarachchige~DCB Gunatilleke, Winnie Wong-Ng, Teiyan Chang, Yu-Sheng Chen, and George~S Nolas.
\newblock Structure, electrical and thermal properties of single-crystal {BaCuGdTe$_3$}.
\newblock {\em Dalton Transactions}, 52(24):8316--8321, 2023.

\bibitem{kresse1996efficiency}
Georg Kresse and J{\"u}rgen Furthm{\"u}ller.
\newblock Efficiency of ab-initio total energy calculations for metals and semiconductors using a plane-wave basis set.
\newblock {\em Computational Materials Science}, 6(1):15--50, 1996.

\bibitem{kresse1996efficient}
Georg Kresse and J{\"u}rgen Furthm{\"u}ller.
\newblock Efficient iterative schemes for ab initio total-energy calculations using a plane-wave basis set.
\newblock {\em Physical Review B}, 54(16):11169, 1996.

\bibitem{kresse1999ultrasoft}
Georg Kresse and Daniel Joubert.
\newblock From ultrasoft pseudopotentials to the projector augmented-wave method.
\newblock {\em Physical Review B}, 59(3):1758, 1999.

\bibitem{setyawan2010high}
Wahyu Setyawan and Stefano Curtarolo.
\newblock High-throughput electronic band structure calculations: Challenges and tools.
\newblock {\em Computational Materials Science}, 49(2):299--312, 2010.

\bibitem{blochl1994projector}
Peter~E Bl{\"o}chl.
\newblock Projector augmented-wave method.
\newblock {\em Physical Review B}, 50(24):17953, 1994.

\bibitem{perdew1996generalized}
John~P Perdew, Kieron Burke, and Matthias Ernzerhof.
\newblock Generalized gradient approximation made simple.
\newblock {\em Physical Review Letters}, 77(18):3865, 1996.

\bibitem{born1926quantenmechanik}
Max Born.
\newblock Quantenmechanik der {Sto{\ss}vorg{\"a}nge}.
\newblock {\em Zeitschrift f{\"u}r Physik}, 38(11):803--827, 1926.

\bibitem{frohlich1954electrons}
Herbert Fr{\"o}hlich.
\newblock Electrons in lattice fields.
\newblock {\em Advances in Physics}, 3(11):325--361, 1954.

\bibitem{deringer2011crystal}
Volker~L Deringer, Andrei~L Tchougr{\'e}eff, and Richard Dronskowski.
\newblock Crystal orbital {Hamilton} population ({COHP}) analysis as projected from plane-wave basis sets.
\newblock {\em The Journal of Physical Chemistry A}, 115(21):5461--5466, 2011.

\bibitem{li2014shengbte}
Wu~Li, Jes{\'u}s Carrete, Nebil~A Katcho, and Natalio Mingo.
\newblock Shengbte: A solver of the {Boltzmann} transport equation for phonons.
\newblock {\em Computer Physics Communications}, 185(6):1747--1758, 2014.

\bibitem{li2012thermal}
Wu~Li, Lucas Lindsay, David~A Broido, Derek~A Stewart, and Natalio Mingo.
\newblock Thermal conductivity of bulk and nanowire {Mg$_2$Si$_x$Sn$_{1- x}$} alloys from first principles.
\newblock {\em Physical Review B—Condensed Matter and Materials Physics}, 86(17):174307, 2012.

\bibitem{togo2015first}
Atsushi Togo and Isao Tanaka.
\newblock First principles phonon calculations in materials science.
\newblock {\em Scripta Materialia}, 108:1--5, 2015.

\bibitem{pailhes2014localization}
St{\'e}phane Pailh{\`e}s, H~Euchner, Valentina~M Giordano, R{\'e}gis Debord, A~Assy, S~Gom{\`e}s, A~Bosak, Denis Machon, S~Paschen, and M~De~Boissieu.
\newblock Localization of propagative phonons in a perfectly crystalline solid.
\newblock {\em Physical Review Letters}, 113(2):025506, 2014.

\bibitem{parker2015benefits}
David~S Parker, Andrew~F May, and David~J Singh.
\newblock Benefits of carrier-pocket anisotropy to thermoelectric performance: The case of p-type agbise 2.
\newblock {\em Physical Review Applied}, 3(6):064003, 2015.

\bibitem{baur1974geometry}
WH~Baur.
\newblock The geometry of polyhedral distortions. predictive relationships for the phosphate group.
\newblock {\em Structural Science}, 30(5):1195--1215, 1974.

\end{thebibliography}

\end{document}